\documentclass[a4paper,12pt]{article}
\usepackage[colorlinks=true,linkcolor=blue,urlcolor=blue,filecolor=black,citecolor=red,pdfstartview=FitV,pdftitle={},pdfsubject={},
pdfkeywords={},pdfpagemode=None,bookmarksopen=true]{hyperref}
\usepackage{graphicx}
\usepackage{epstopdf}%
\usepackage{amsmath}
\usepackage{amsfonts}
\usepackage{amssymb}
\usepackage{color}%
\usepackage{dcolumn}
\usepackage{slashed}
\usepackage{amssymb,ulem}
\usepackage{float}
\usepackage{amsthm,amsmath,amssymb}
\usepackage{mathrsfs}
\usepackage{subfigure}
\usepackage{wrapfig}
\usepackage{indentfirst}
\usepackage{multirow}
\usepackage{appendix}
\usepackage{graphicx} 
\usepackage{subfigure} 
\usepackage{tablefootnote}
\usepackage{suffix}
\WithSuffix\newcommand\footnote*[1]{\footnote[#1]{\addtocounter{footnote}{-1}}}

\makeatletter
\renewcommand\@makefnmark{\hbox{\@textsuperscript{\normalfont\@thefnmark}}}

\newcommand{\Rmnum}[1]{\expandafter\@slowromancap\romannumeral #1@}

\makeatother

\begin{document}

\vspace{9mm}

\begin{center}
{{{\Large {\bf Thermodynamics and phase transitions of nonlinearly scalarized black holes in Einstein-scalar-Gauss-Bonnet theory }}}}\\[8mm]

{De-Cheng Zou$^{1}$\footnote{dczou@jxnu.edu.cn;}, Xu Yang$^{1}$\footnote{xuyangjxnu0329@163.com;}, Meng-Yun Lai$^{1}$\footnote{ mengyunlai@jxnu.edu.cn},  Hyat Huang$^{1}$\footnote{hyat@mail.bnu.edu.cn} and Yun Soo Myung$^{2}$\footnote{ysmyung@inje.ac.kr} }\\[8mm]

{
${}^1$School of Physics, Jiangxi Normal University, Nanchang 330022, China\\
$^{2}$Center for Quantum Spacetime, Sogang University, Seoul 04107, Republic of  Korea}

\end{center}

\begin{abstract}
We investigate the thermodynamic properties of static nonlinearly scalarized black holes in Einstein-scalar-Gauss-Bonnet theory with polynomial coupling functions. Based on the scalarized solutions constructed previously, we compute thermodynamical quantities of these scalarized black holes. Moreover, we examine the first law of black hole thermodynamics and consider the phase transitions between Schwarzschild and scalarized black holes. It shows that a phase transition from Schwarzschild black hole to scalarized black hole is a first-order with non-zero latent heat. 
\end{abstract}


\section{Introduction}\label{1s}

Scalar fields are a fundamental and widespread feature in various branches of theoretical physics, providing an elegant and adaptable framework for modeling the accelerated expansion of the universe, from its primordial epochs to its present state. However, the no-hair theorem rules out a black hole with scalar hair in the Einstein-minimally coupled scalar theory~\cite{Bekenstein:1974sf}-\cite{Bekenstein:1995un}. This theorem can be circumvented when the theory possesses a nonminimal scalar coupling to the curvature/matter. In this direction, Damour and Esposito-Farese~\cite{Damour:1993hw} have shown a mechanism of spontaneous scalarization in scalar-tensor theory  when studying neutron stars.
Over the past several years, Einstein-scalar-Gauss-Bonnet (EsGB) gravity has been widely studied in connection with black hole spontaneous scalarization. In this theory, a dynamical scalar field couples non-minimally to the Gauss–Bonnet invariant, endowing black holes with scalar hair while preserving second-order field equations in four dimensions~\cite{Doneva:2017bvd}-\cite{Antoniou:2017acq}. 
For both Schwarzschild~\cite{Doneva:2017bvd}-
\cite{Blazquez-Salcedo:2024rvb} and Kerr~\cite{Dima:2020yac}-\cite{Zou:2021ybk} black holes, they become unstable at the linear level as a scalar perturbation acquires an effective tachyonic mass in a region of sufficiently strong curvature. 
The corresponding scalarized branches usually bifurcate from the bald(Schwarzschild/Kerr) solutions at the onset of the linear instability.

Note that scalarization may also occur without a linear tachyonic instability. For coupling functions satisfying \( \zeta''(0)=0 \), Ref.~\cite{Doneva:2021tvn} recovered that the effective mass of the scalar perturbation around the Schwarzschild black hole vanishes, and the Schwarzschild black hole remains linearly stable. Even so, scalar hair can still be triggered by finite-amplitude perturbations in EsGB theory for exponential couplings. This mechanism is known as \textbf{nonlinear scalarization}. Then, it was studied from several aspects, including radial stability, self-interaction effects, and rotating black holes~\cite{Blazquez-Salcedo:2022omw}-\cite{Xiong:2023bpl}.
Recently, nonlinear scalarization was further studied for polynomial coupling functions in EsGB theory~\cite{Zou:2024wsk} . In that work, threshold amplitudes for the onset of scalarization were identified from the time evolution of scalar perturbations on a Schwarzschild background, and fully backreacted static scalarized black holes were constructed. The resulting solutions show a nontrivial branch structure, including disconnected branches and branch mergers in some parameter regions.

On the other hand, the thermodynamic aspects of scalarized black holes in EsGB gravity have also received increasing attention. In particular, the phase structure of scalarized black holes was recently analyzed for several classes of scalar-Gauss-Bonnet couplings, showing that the thermodynamic behavior depends sensitively on the form of the coupling function~\cite{Herdeiro:2026sur}. This makes it natural to examine the thermodynamic properties of the polynomial-coupling solutions constructed in ~\cite{Zou:2024wsk}.
In this paper, we plan to investigate the thermodynamic properties of nonlinearly scalarized black holes for polynomial coupling cases~\cite{Zou:2024wsk}. We will further check the first law of thermodynamics, and discuss phase transitions by comparing the scalarized black holes with the Schwarzschild black holes.

This paper is organized as follows. In Sect.~\ref{Sec2}, we briefly review the EsGB model with polynomial couplings and the corresponding scalarized black hole solutions. In Sect.~\ref{Sec3}, we present thermodynamic quantities and discuss the phase structure of the different solution branches. Finally, Sect.~\ref{Sec4} contains the conclusions and discussion.

\section{EsGB Model and Scalarized Black Hole Solutions}
\label{Sec2}

We consider the Einstein-scalar-Gauss-Bonnet (EsGB) theory with action
\begin{eqnarray}
S=\frac{1}{16\pi}\int d^4x\sqrt{-g}\left[R-2\nabla_\mu\phi\nabla^\mu\phi+\lambda^2\zeta(\phi)R^2_{\rm GB}\right],
\end{eqnarray}
where
\begin{eqnarray}
R^2_{\rm GB}=R_{\mu\nu\rho\sigma}R^{\mu\nu\rho\sigma}-4R_{\mu\nu}R^{\mu\nu}+R^2
\end{eqnarray}
is the Gauss-Bonnet invariant. The field equations are
\begin{eqnarray}
R_{\mu\nu}-\frac{1}{2}Rg_{\mu\nu}+\Gamma_{\mu\nu}
=2\nabla_\mu\phi\nabla_\nu\phi-g_{\mu\nu}\nabla_\alpha\phi\nabla^\alpha\phi,
\end{eqnarray}
\begin{eqnarray}
\square\phi+\frac{\lambda^2}{4}\frac{d\zeta(\phi)}{d\phi}R^2_{\rm GB}=0,
\end{eqnarray}
where
\begin{eqnarray}
\Gamma_{\mu\nu}&=&-R(\nabla_\mu\psi_\nu+\nabla_\nu\psi_\mu)
-4\nabla_\alpha\psi^\alpha\left(R_{\mu\nu}-\frac{1}{2}Rg_{\mu\nu}\right)
+4R_{\mu\alpha}\nabla^\alpha\psi_\nu \nonumber\\
&&+4R_{\nu\alpha}\nabla^\alpha\psi_\mu
-4g_{\mu\nu}R_{\alpha\beta}\nabla^\alpha\psi^\beta
+4R_{\beta\mu\alpha\nu}\nabla^\alpha\psi^\beta,
\end{eqnarray}
with \(\psi_\mu=\lambda^2 \frac{d\zeta(\phi)}{d\phi}\nabla_\mu\phi\).

For static scalarized black holes, we adopt the spherically symmetric ansatz
\begin{eqnarray}
ds^2=-A(r)dt^2+\frac{dr^2}{B(r)}+r^2d\Omega_2^2.
\end{eqnarray}
The numerical solutions are constructed by imposing a regular event horizon at \(r=r_H\), with \(A(r_H)=B(r_H)=0\), together with asymptotic flatness at spatial infinity, where \(A(r)\to1\), \(B(r)\to1\), and \(\phi(r)\to0\).
Regularity of the scalar field at the horizon further requires
\begin{eqnarray}
\Delta\equiv 1-\frac{24\lambda^4}{r_H^4}\left(\frac{d\zeta}{d\phi}(\phi_H)\right)^2\ge0,
\end{eqnarray}
where \(\phi_H\) denotes the scalar field at the horizon. We also monitor the Kretschmann scalar in order to exclude solutions with curvature singularities outside the horizon.

In this work, we consider the polynomial couplings
\begin{eqnarray}
\zeta_1(\phi)=\alpha\phi^4-\beta\phi^8,\quad
\zeta_2(\phi)=\alpha\phi^4-\beta\phi^6,\quad
\zeta_3(\phi)=\alpha\phi^4.
\end{eqnarray}
For \(\zeta_1(\phi)\) and \(\zeta_2(\phi)\), the basic relations \(\phi_H(M)\) and \(Q_s(M)\) are qualitatively similar~\cite{Zou:2024wsk}. In the following, we focus on \(\zeta_1(\phi)\) firstly.

A characteristic feature of the polynomial couplings is the appearance of multiple scalarized branches. For \(\zeta_1(\phi)=\alpha\phi^4-\beta\phi^8\), keeping \(\Delta\) finite as \(M\to0\) requires
\begin{eqnarray}
\frac{d\zeta_1}{d\phi}(\phi_H)\sim0 \Rightarrow
\phi_H^3(4\alpha-8\beta\phi_H^4)\sim0.
\end{eqnarray}
Besides the trivial root \(\phi_H\sim0\), this equation admits another real root,
\begin{eqnarray}
\phi_H\sim\left(\frac{\alpha}{2\beta}\right)^{1/4},
\end{eqnarray}
which explains the appearance of additional branches. By contrast, for the quartic coupling \(\zeta_3(\phi)=\alpha\phi^4\), only the root \(\phi_H=0\) remains.

\begin{figure}[H]
\centering
\subfigure[$\phi_H$ vs $ M$]{
\label{3a} 
\includegraphics[width=0.3\textwidth]{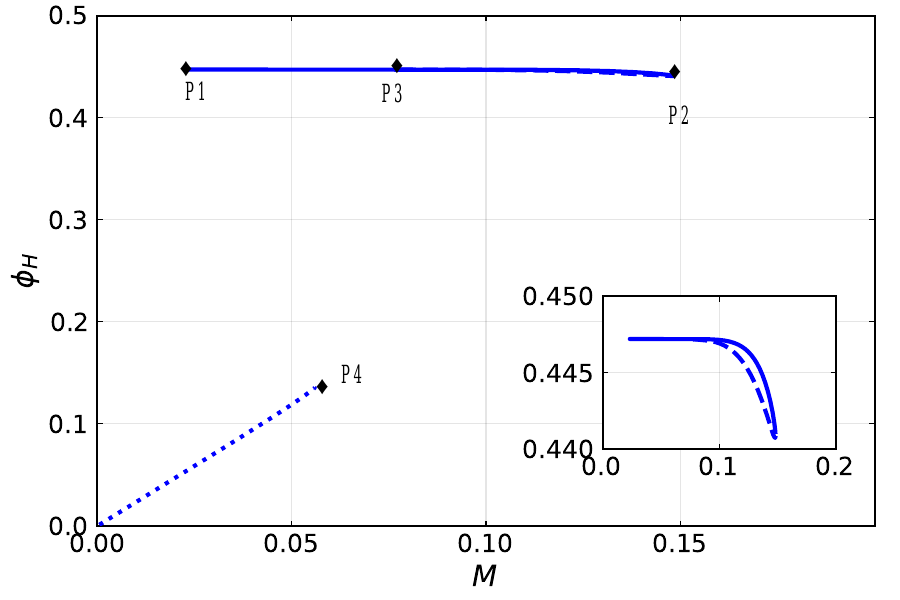}}
\hfill
\subfigure[$Q_s$ vs $ M$]{
\label{3b} 
\includegraphics[width=0.3\textwidth]{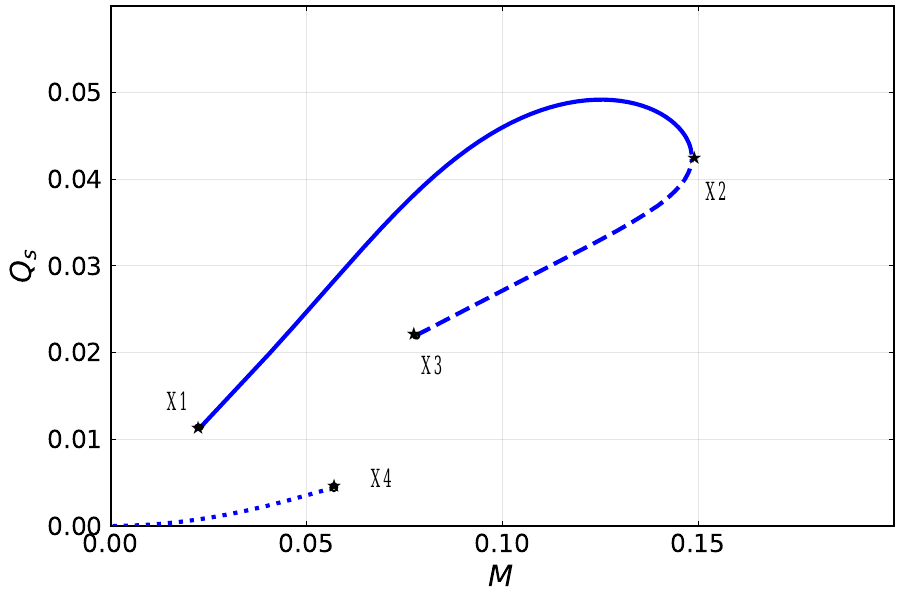}}
\hfill
\subfigure[$\Delta$ vs $ M$]{
\label{3c} 
\includegraphics[width=0.3\textwidth]{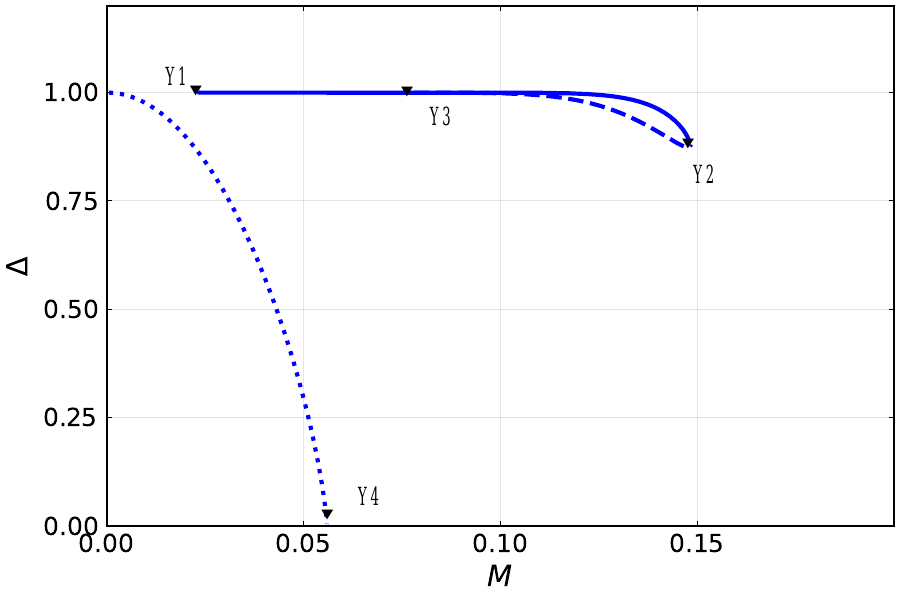}}
\caption{Scalar hair $\phi_H$ at the horizon, scalar charge $Q_s$, and  regularity parameter  $\Delta$ as  functions of the mass $M$ for coupling function $\zeta_1(\phi)$ with $\alpha=1/4$ and $\beta=25/8$. We find three branches (solid, dashed, dotted) of scalarized solutions}\label{fig1}
\end{figure}

\begin{figure}[H]
\centering
\subfigure[$\phi_H$ vs $ M$]{
\label{6a} 
\includegraphics[width=0.3\textwidth]{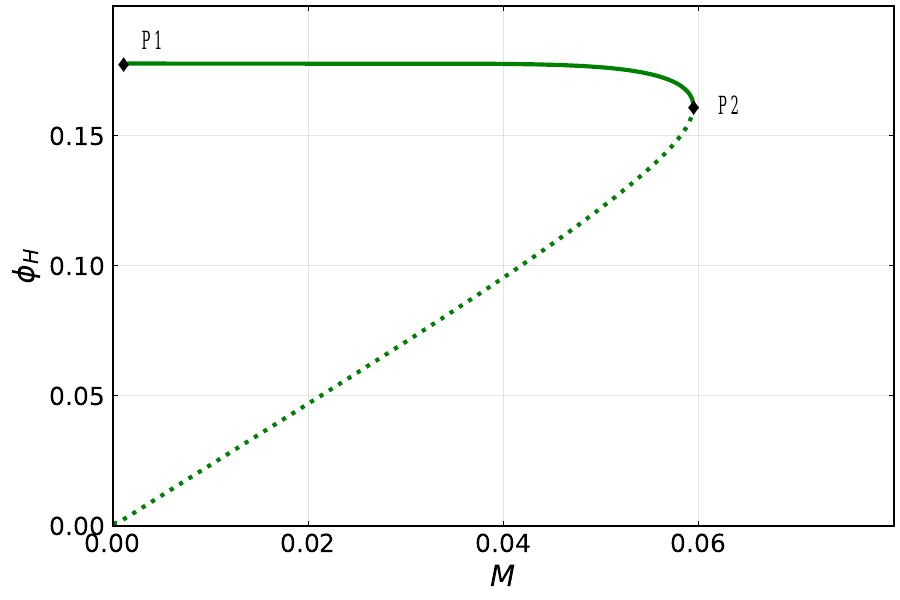}}
\hfill%
\subfigure[$Q_s$ vs $ M$]{
\label{6b} 
\includegraphics[width=0.3\textwidth]{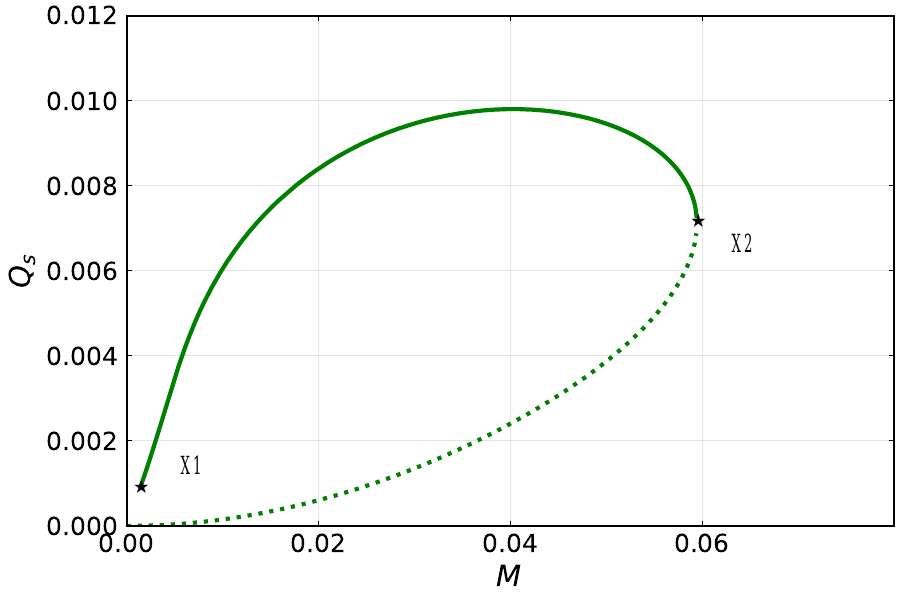}}%
\hfill%
\subfigure[$\Delta$ vs $ M$]{
\label{6c} 
\includegraphics[width=0.3\textwidth]{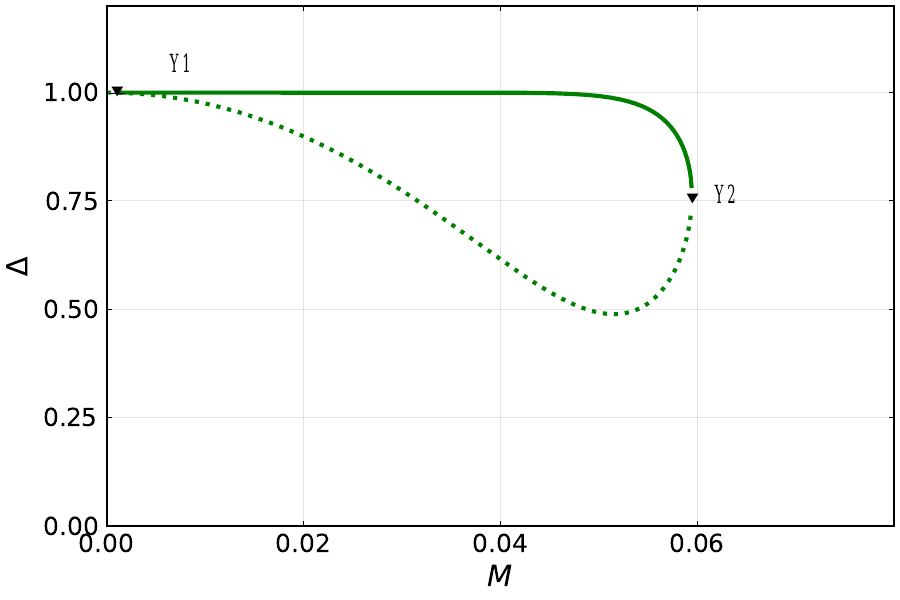}}
\caption{Scalar hair $\phi_H$  at the horizon, scalar charge $Q_s$ and  regularity parameter  $\Delta$ as a function of mass $M$ for coupling function $\zeta_1(\phi)$ with $\alpha=1/4$ and $\beta=1000/8$. 
}\label{fig2}
\end{figure}

The detailed branch structure for different values of \(\beta\) has already been presented in Ref.~\cite{Zou:2024wsk}. Here we only show two representative cases. For small \(\beta\), three scalarized branches are found, as shown in Fig.~\ref{fig1} for \(\beta=25/8\). One branch is connected to the Schwarzschild limit, while the other two form a disconnected pair that merge at a turning point. The allowed mass interval is bounded either by \(\Delta=0\) or by the appearance of a curvature singularity outside the horizon, as indicated by the Kretschmann scalar~\cite{Zou:2024wsk}. For large \(\beta\), the branch structure becomes simpler: as shown in Fig.~\ref{fig2} for \(\beta=1000/8\), only two branches remain and merge at a point.

\begin{figure}[H]
\centering
\subfigure[ $A(r)$, $B(r)$ and $f(r)$ as  functions of $r$]{
\label{8a} 
\includegraphics[width=0.3\textwidth]{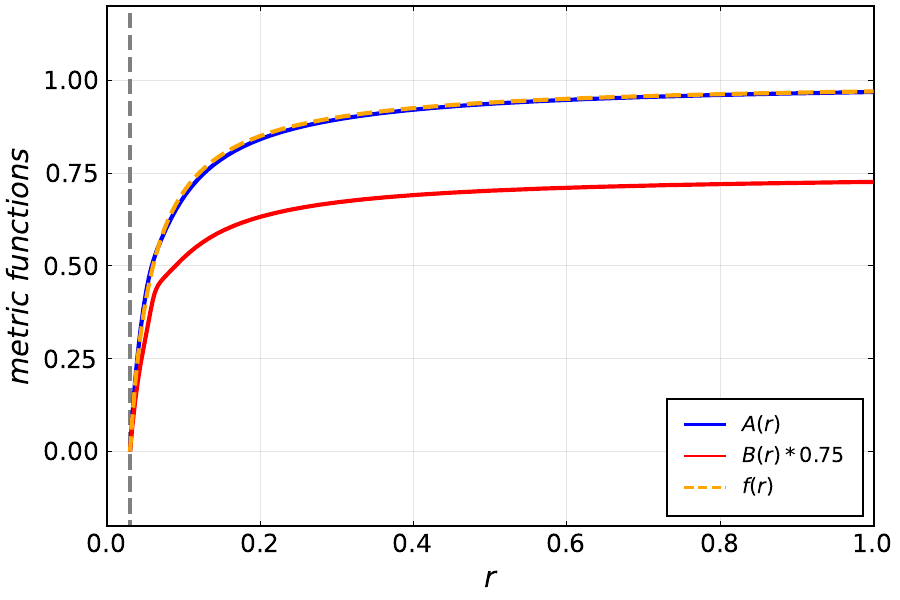}}
\hfill%
\subfigure[ $\delta A(r)$ and $\delta B(r)$ as  functions of $r$]{
\label{8b}
\includegraphics[width=0.3\textwidth]{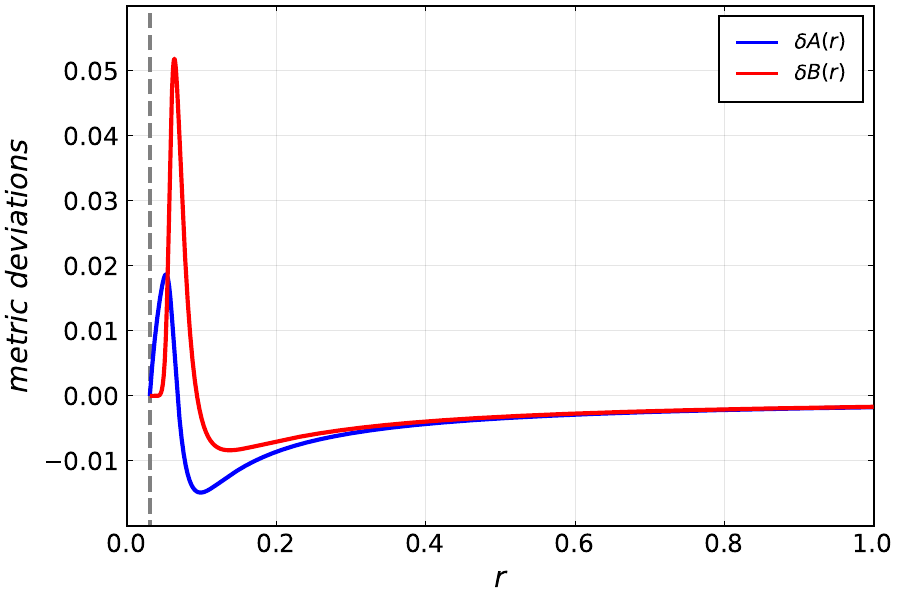}}
\hfill%
\subfigure[Scalar hair  $\phi(r)$ as functions of $r$]{
\label{8c}
\includegraphics[width=0.3\textwidth]{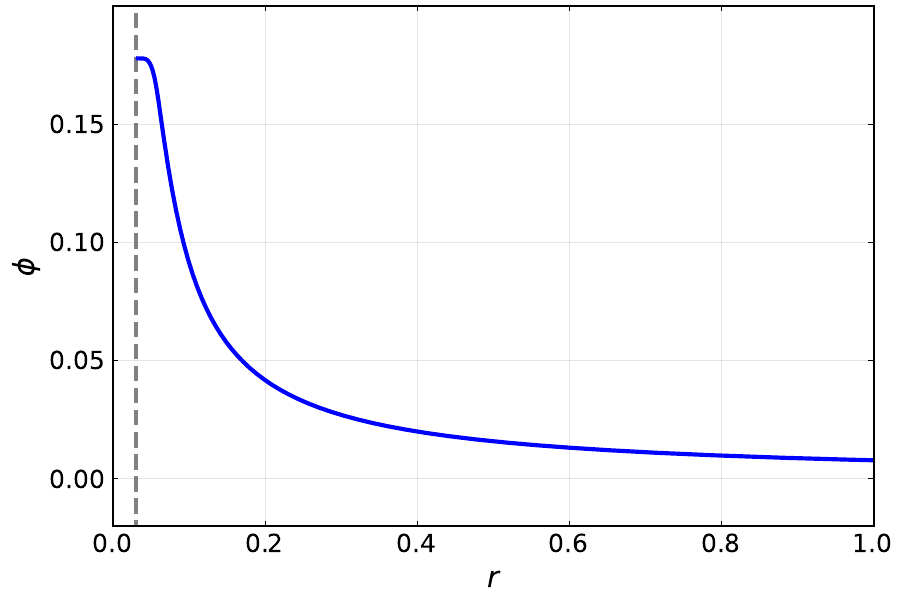}}
\caption{Graph for scalarized BH  with $r_H= 0.03$. Here, the coupling function is $\zeta_1(\phi)$ with $\alpha=1/4$ and $\beta = 1000/8$.}\label{fig3}
\end{figure}

A representative scalarized solution for \(\zeta_1(\phi)\) is shown in Fig.~\ref{fig3}. The metric functions are regular outside the horizon and approach the Schwarzschild form at large radius, while the scalar field decreases monotonically and vanishes asymptotically.
In addition, the simpler quartic case \(\zeta_3(\phi)=\alpha\phi^4\) is summarized in Appendix~A.

\section{Thermodynamics and Phase Transition}
\label{Sec3}

In this section, we study the thermodynamic properties of the nonlinearly scalarized black holes in EsGB theory. Based on the numerical solutions obtained above, we compute the thermodynamical quantities of scalarized black holes.

The Hawking temperature is given by
\begin{eqnarray}
T_H=\frac{1}{4\pi}\sqrt{A'(r_H)B'(r_H)}.
\end{eqnarray}
The Wald entropy of the scalarized black holes in the EsGB theory takes the form~\cite{Doneva:2017bvd,Antoniou:2017hxj,Gibbons:1976ue}
\begin{eqnarray}
S=\frac{1}{4}A_H+4\pi \lambda^2 \zeta(\phi_H),
\end{eqnarray}
where \(A_H=4\pi r_H^2\) is the horizon area.

\begin{figure}[H]
\centering
\subfigure[$\beta=25/8$]{\label{10a}\includegraphics[width=0.4\textwidth]{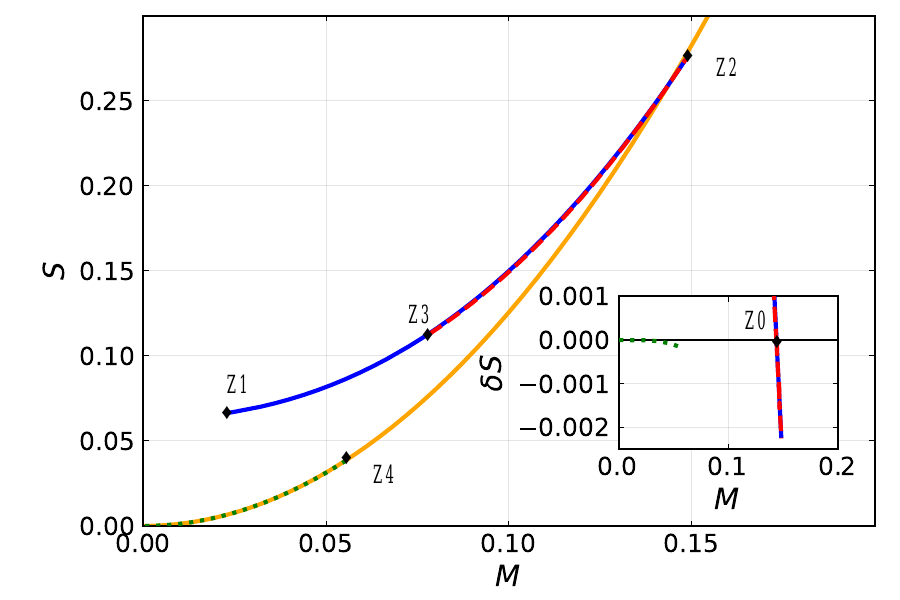}}
\hfill%
\subfigure[$\beta=278/80$]{\label{10b}\includegraphics[width=0.4\textwidth]{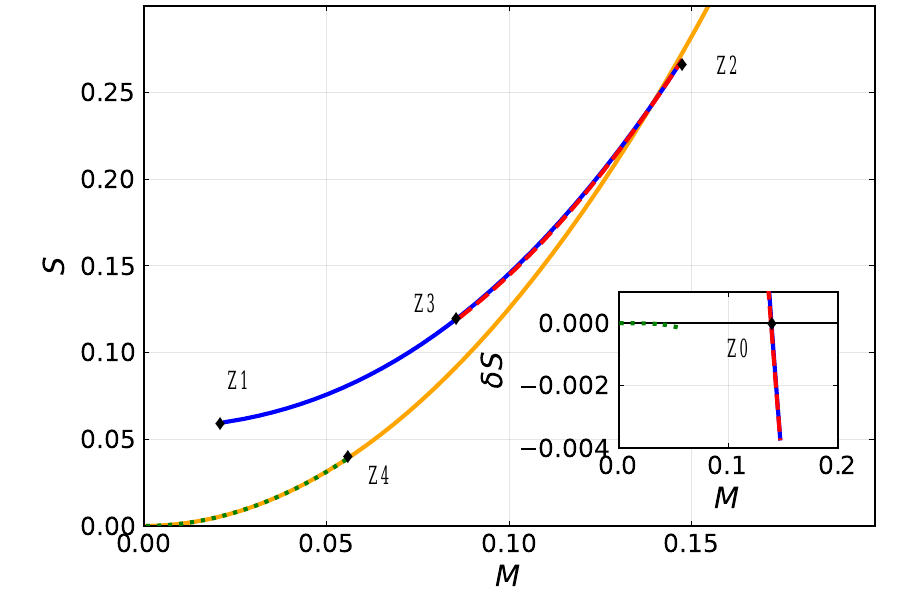}}
\hfill%
\subfigure[$\beta=100/8$]{\label{10c}
\includegraphics[width=0.4\textwidth]{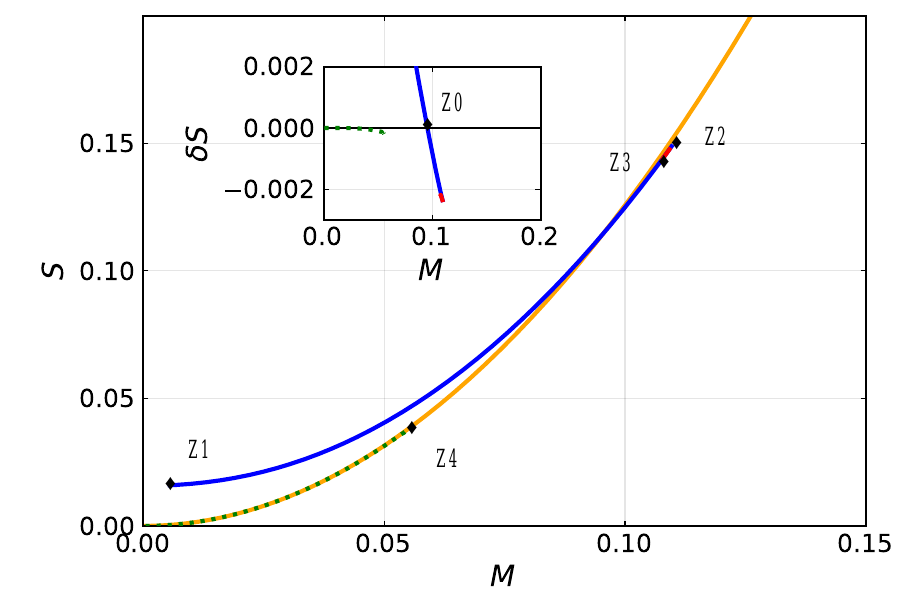}}%
\hfill%
\subfigure[$\beta=1000/8$]{\label{10d}
\includegraphics[width=0.4\textwidth]{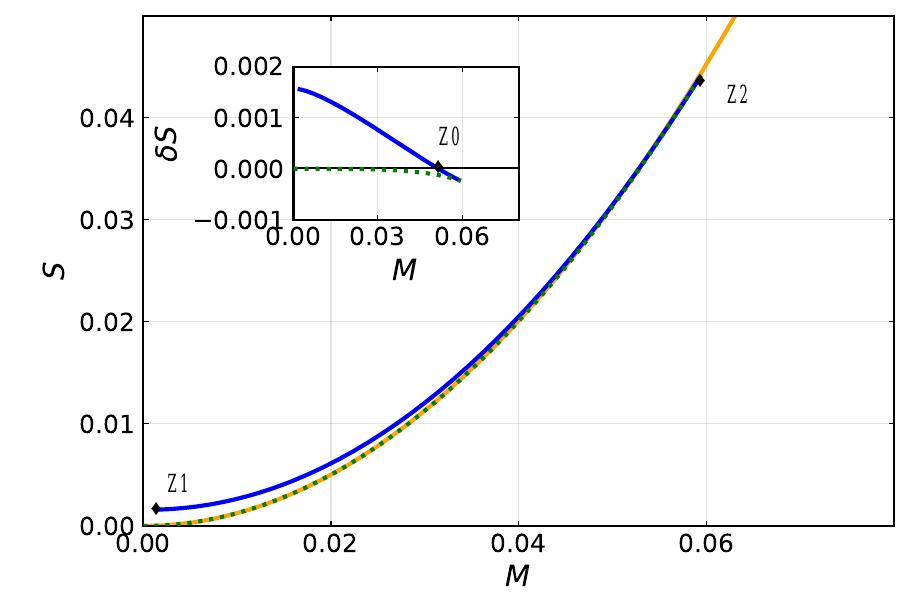}}%
\caption{The entropy $S$ as a function of
mass $M$ for polynomial coupling function $\zeta_1(\phi)$ with four different $\beta$. Here, $\delta S=S-S_0$ with $S_0$ for Schwarzschild black hole.}\label{fig4}
\end{figure}

We focus on the model with coupling
\begin{eqnarray}
\zeta_1(\phi)=\frac{\phi^4}{4}-\beta \phi^8,
\end{eqnarray}
where \(\beta\) serves as the control parameter. Fig.~\ref{fig4} shows the entropy \(S\) as a function of the black hole mass \(M\) for several values of \(\beta\). Along all branches, the entropy increases monotonically with the mass. To compare the scalarized and Schwarzschild solutions, we introduce the entropy difference
\begin{eqnarray}
\delta S=S-S_0,
\end{eqnarray}
where \(S_0\) denotes the entropy of the Schwarzschild black hole with the same mass. The condition \(\delta S=0\) defines a critical mass \(M_c\). For \(M<M_c\), one has \(\delta S<0\), so the Schwarzschild black hole is entropically favored. For \(M\ge M_c\), one has \(\delta S>0\), and the scalarized black hole has larger entropy.

\begin{figure}[H]
\centering
\subfigure[$\beta=25/8$]{\label{11a}
\includegraphics[width=0.4\textwidth]{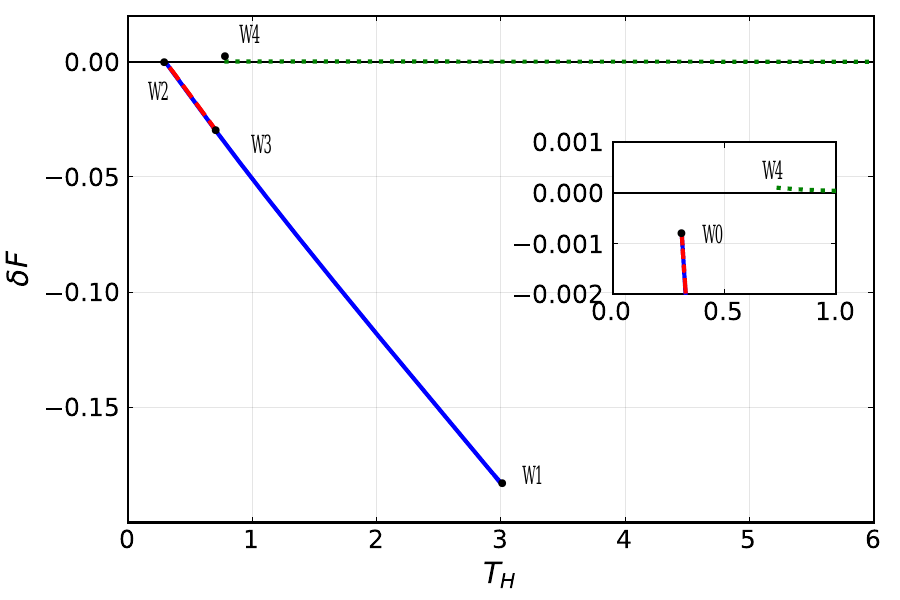}}%
\hfill%
\subfigure[$\beta=278/80$]{\label{11b}
\includegraphics[width=0.4\textwidth]{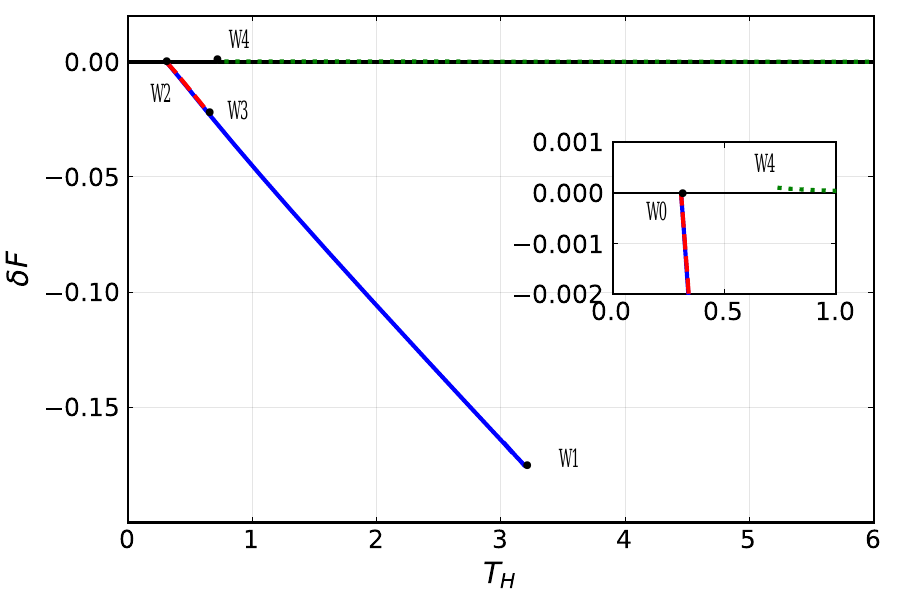}}
\hfill%
\subfigure[$\beta=100/8$]{\label{11c}
\includegraphics[width=0.4\textwidth]{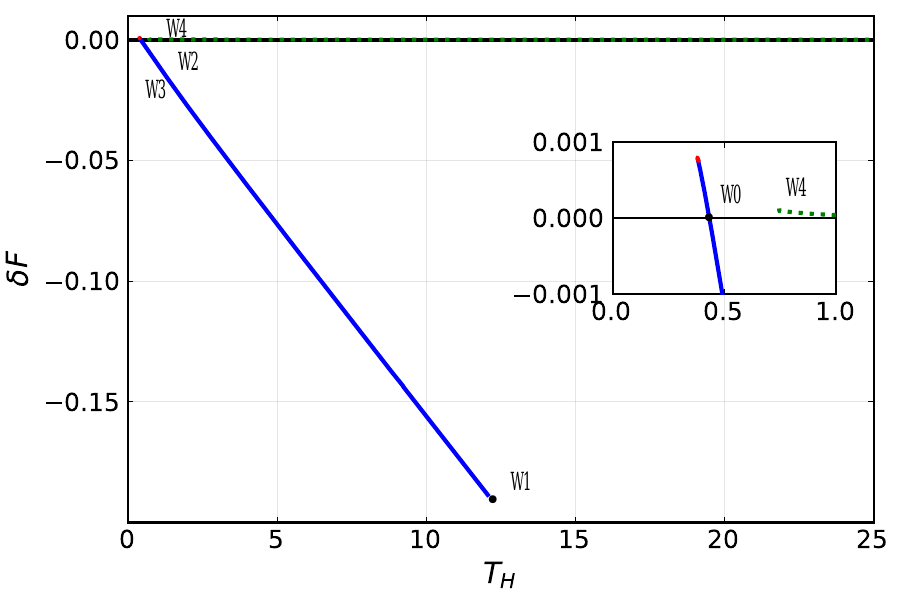}}
\hfill%
\subfigure[$\beta=1000/8$]{\label{11d}
\includegraphics[width=0.4\textwidth]{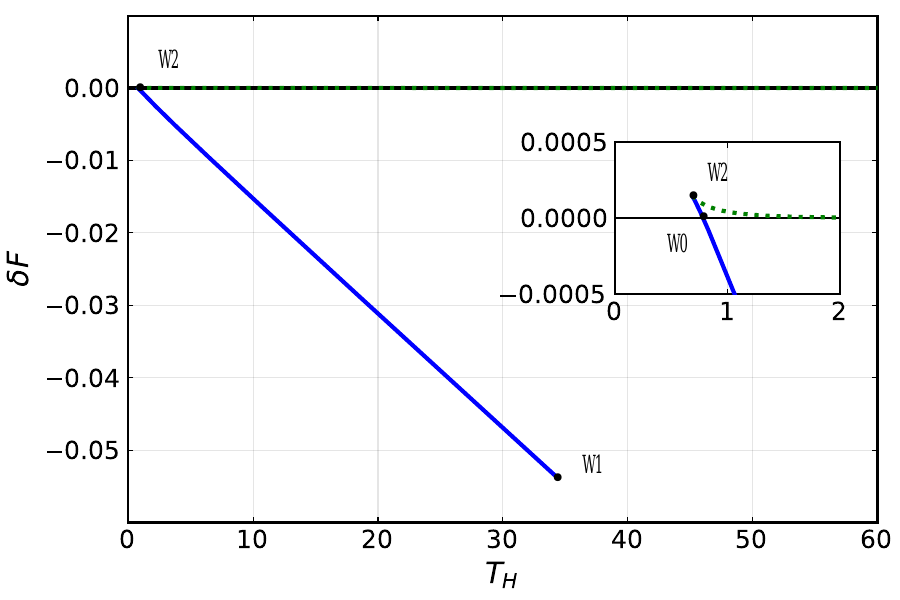}}
\caption{ 
Difference of free energy $\delta F=F_{\rm scal}-F_{\rm SBH}$ as a function of Hawking temperature  $T_H$ for $\zeta_1(\phi)$ with four different $\beta$.}\label{fig5}
\end{figure}

The global thermodynamic preference is determined by the Helmholtz free energy. In Fig.~\ref{fig5}, we plot the free-energy difference
\begin{eqnarray}
\delta F=F_{\rm scal}-F_{\rm SBH}
\end{eqnarray}
as a function of the Hawking temperature \(T_H\). A negative value of \(\delta F\) means that the scalarized black hole is thermodynamically favored over the Schwarzschild black hole. For a given \(\beta\), \(\delta F\) changes sign at a critical temperature \(T_c\), marked by the crossing point where \(\delta F=0\). This identifies a phase transition between the Schwarzschild and scalarized phases.

A key point is that the conditions \(\delta S=0\) and \(\delta F=0\) do not occur at the same state. Therefore, the phase transition is not determined simply by the entropy comparison, but by the free-energy competition between the two phases. At \(T=T_c\), the two phases have the same Helmholtz free energy but different entropies, which implies a finite latent heat,
\begin{eqnarray}
L=T_c\,\Delta S.
\end{eqnarray}
This shows that the transition from the Schwarzschild black hole to the scalarized black hole is of first order.

\begin{figure}[H]
\centering
\subfigure[$\delta S$ vs $ M$]{
\includegraphics[width=0.45\textwidth]{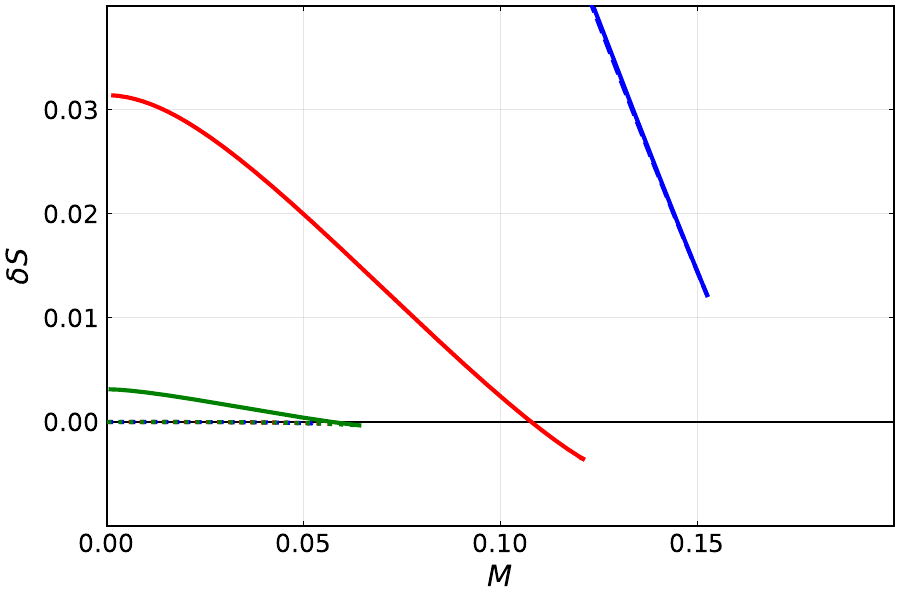}}%
\hfill%
\subfigure[$\delta F$ vs $ T_H$]{
\includegraphics[width=0.45\textwidth]{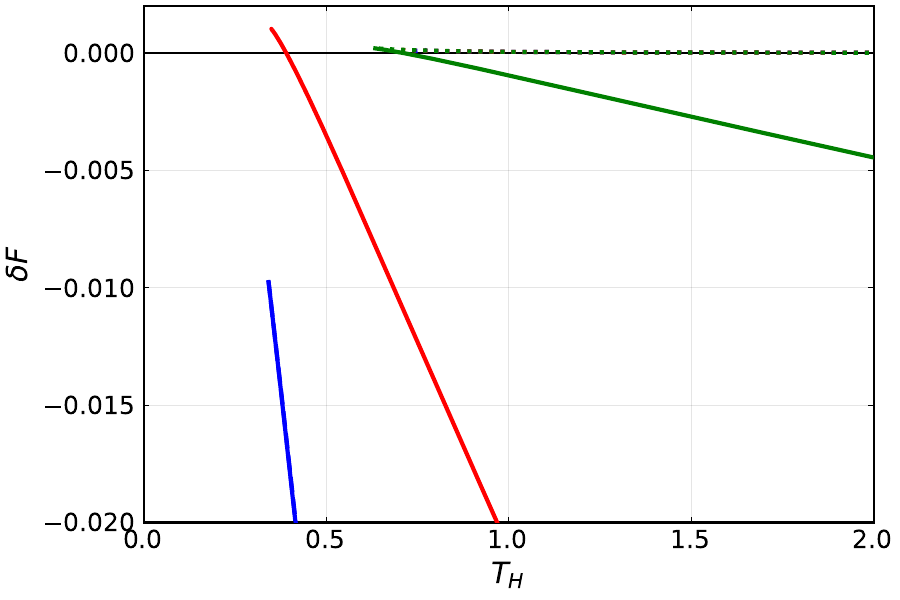}}%
\caption{ 
Two differences of $\delta S$  and $\delta F$ for exponential coupling function $\zeta_e(\phi)=\frac{1}{4\beta}\left(1-e^{-\beta\phi^4}\right)$ with three different curves  of $\beta=25$ (blue), 100(red) and 1000(green).}\label{fig6}
\end{figure}

The parameter \(\beta\) plays an important role in the thermodynamic behavior of the system. As \(\beta\) increases, the free-energy landscape becomes smoother and the latent heat decreases significantly, as shown in Table~\ref{Table1}. At the same time, the thermodynamically favored scalarized phase is shifted to higher temperatures. Similar behavior is also found for the exponential coupling \(\zeta_e(\phi)\), as illustrated in Fig.~\ref{fig6} and Table~\ref{Table1}.

\begin{table*}[h!]
\caption{The latent heat of first order phase transition between scalarized BHs and SBHs for polynomial $(L_p)$ and exponential $(L_e)$ couplings.}\label{Table1}
\centering
\begin{tabular}{|c|c|c|} \hline
  $\beta$   &  $L_p=T_c|\delta S|$ &  $L_e=T_c|\delta S|$ \\ \hline
100/8&   $2.773\times10^{-4}$ &  $6.064\times10^{-4}$     \\ 
1000/8 &  $1.658\times10^{-5}$  & $3.744\times10^{-5}$     \\ 
2000/8 &  $7.020\times10^{-6}$  & $1.590\times10^{-5}$     \\
3000/8 &  $4.242\times10^{-6}$  & $9.624\times10^{-6}$     \\
6000/8 & $1.791\times10^{-6}$ &  $4.070\times10^{-6}$ \\
10000/8 & $9.478\times10^{-7}$ & $2.156\times10^{-6}$  \\ \hline
\end{tabular}
\end{table*}

For comparison, we also consider the quartic coupling
\begin{eqnarray}
\zeta_3(\phi)=\alpha\phi^4.
\end{eqnarray}
In this case, although regular scalarized black hole solutions exist, their free energy is never lower than that of the Schwarzschild black hole, i.e. \(\delta F\ge0\) over the whole temperature range. Therefore, no thermodynamic phase transition is expected for the quartic coupling. The corresponding numerical results are summarized in Appendix~A.

\begin{figure}[ht]
\centering
\includegraphics[width=0.5\textwidth]{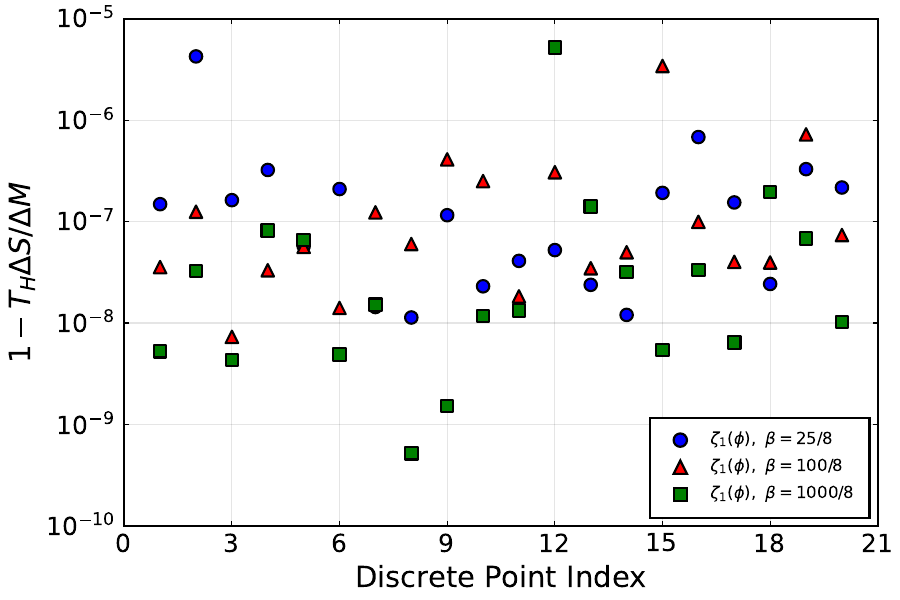}
\caption{Twenty  random solutions  are distributed for each branch to check the first-law of thermodynamics.}\label{fig7}
\end{figure}

Finally, we test the first law of black hole thermodynamics numerically. Using the discrete solution data, we define the finite differences
\begin{eqnarray}
\Delta M=\frac{M(i+2)-M(i)}{2}, \qquad
\Delta S=\frac{S(i+2)-S(i)}{2},
\end{eqnarray}
and evaluate the quantity
\begin{eqnarray}
1-\frac{T_H\Delta S}{\Delta M}.
\end{eqnarray}
If the first law \(dM=T_H dS\) is exactly satisfied, this quantity should vanish. The numerical results are shown in Fig.~\ref{fig7} for several values of \(\beta\). The deviations are typically of order \(10^{-5}\sim 10^{-10}\), indicating that the first law is satisfied with good numerical accuracy.

\section{Conclusions and discussions}
\label{Sec4}

In this work, we studied the thermodynamic properties and phase transitions of nonlinearly scalarized black holes in Einstein-scalar-Gauss-Bonnet theory. Since the solutions for \(\zeta_2(\phi)\) are qualitatively similar to those for \(\zeta_1(\phi)\), we focused mainly on the branch structure and thermodynamics of the \(\zeta_1(\phi)\) case by solving the full field equations. For \(\alpha=1/4\), the number of scalarized branches depends on the control parameter \(\beta\): three branches are found for \(\beta=25/8\) and \(100/8\), while two branches remain for \(\beta=1000/8\). We also exhibited scalarized solutions close to the Schwarzschild black hole for both \(\zeta_1(\phi)\) and \(\zeta_2(\phi)\).

Using these solutions, we computed the Hawking temperature \(T_H\), entropy \(S\), and Helmholtz free energy \(F\) for the polynomial coupling $\zeta_1(\phi)=\frac{\phi^4}{4}-\beta\phi^8$,
with \(\beta\) as the control parameter. We also checked the first law numerically. The entropy difference \(\delta S=S-S_0\), where \(S_0\) is the entropy of the Schwarzschild black hole with the same mass, defines a critical mass \(M_c\) through \(\delta S=0\). For \(M<M_c\), the Schwarzschild black hole is entropically favored, whereas for \(M\ge M_c\), the scalarized black hole has larger entropy.
The global thermodynamic preference is determined by the free-energy difference $\delta F=F_{\rm scal}-F_{\rm SBH}$.
For a given \(\beta\), \(\delta F\) changes sign at a critical temperature \(T_c\). Since this crossing is accompanied by a discontinuous jump in entropy, the transition from the Schwarzschild black hole to the scalarized black hole is of first order and carries a nonzero latent heat. Our results also show that \(\beta\) strongly affects the thermodynamic behavior: as \(\beta\) increases, the latent heat decreases and the stable scalarized phase is shifted to higher temperatures. The fact that the conditions \(\delta S=0\) and \(\delta F=0\) do not coincide further supports the first-order nature of the transition.

For comparison, we also examined the quartic coupling \(\zeta_3(\phi)=\alpha\phi^4\). Although regular scalarized black hole solutions exist in this case, their free energy is never lower than that of the Schwarzschild black hole. Therefore, no thermodynamic phase transition is expected, and the scalarized solutions should instead be regarded as metastable configurations.

Overall, our results show that nonlinearly scalarized black holes in EsGB theory can exhibit a clear thermodynamic phase structure. In the polynomial-coupling case studied here, the transition from the Schwarzschild phase to the scalarized phase is first order with nonzero latent heat.

\vspace{1cm}

{\bf Acknowledgments}

\vspace{1cm}

We gratefully acknowledge support by the National Natural Science Foundation of China (NNSFC) (Grant Nos.12365009, 12305064 and 12565010). Y.S.M. was supported by the National Research Foundation of Korea (NRF) grant
 funded by the Korea government(MSIT) (RS-2022-NR069013).

 \vspace{1cm}

\appendix
\section{Computation for \texorpdfstring{$\zeta_3(\phi)=\alpha\phi^4$}{zeta3(phi)=alpha phi4}}

For comparison with the polynomial couplings \(\zeta_1(\phi)\) and \(\zeta_2(\phi)\), we also briefly examine the quartic coupling
\[
\zeta_3(\phi)=\alpha\phi^4.
\]

\begin{figure}[ht]
\centering
\subfigure[$Q_s$ vs $ M$]{
\includegraphics[width=0.4\textwidth]{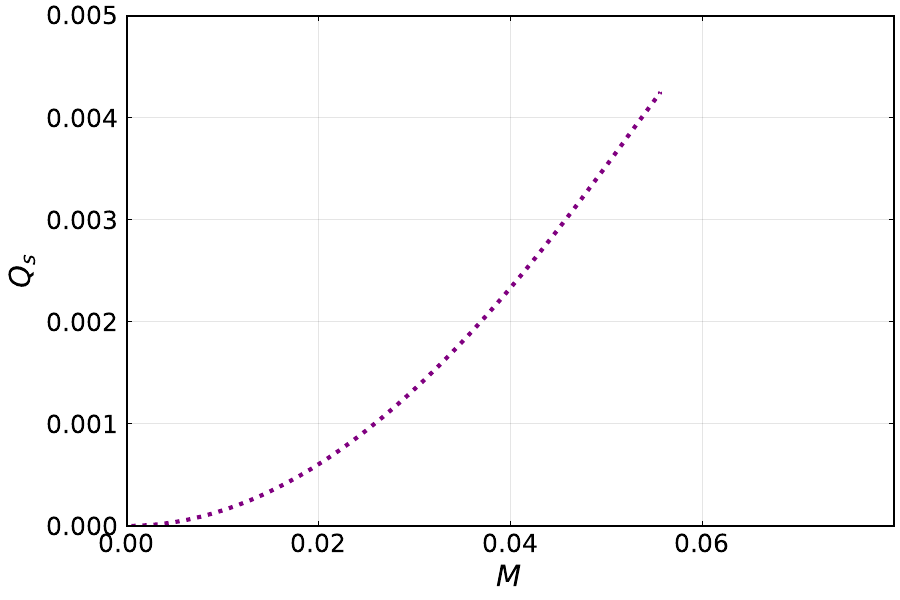}}
\hfill%
\subfigure[$\Delta$ vs $ M$]{
\includegraphics[width=0.4\textwidth]{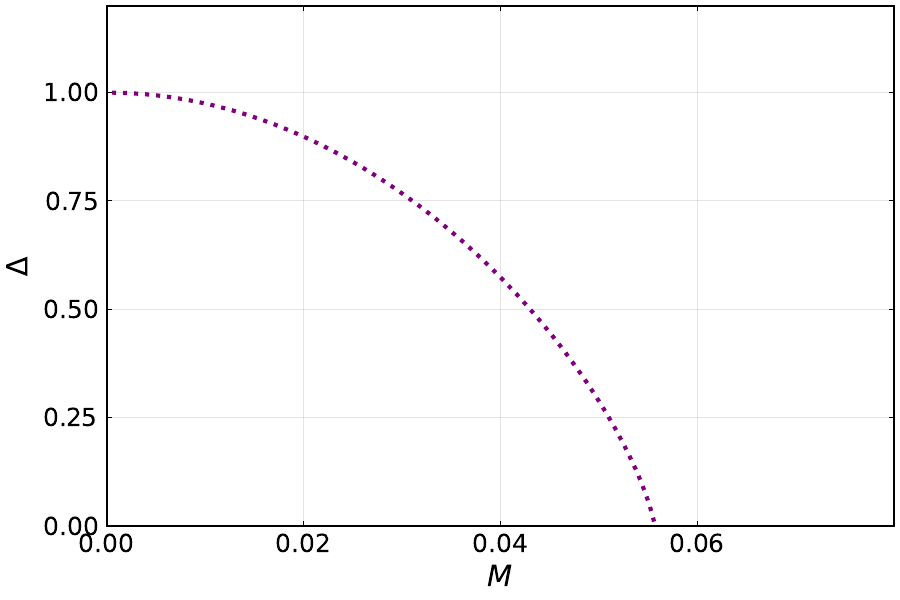}}
\caption{Scalar charge $Q_s$ and  regularity  parameter $\Delta$ with $\zeta_3(\phi)=\alpha\phi^4$.}\label{fig8}
\end{figure}

The existence and basic properties of the corresponding static scalarized black holes are shown in Fig.~\ref{fig8}. In Fig.~\ref{fig8}(a), the scalar charge \(Q_s\) increases monotonically with the black hole mass \(M\), indicating the existence of a continuous family of scalarized solutions. Figure~\ref{fig8}(b) shows the regularity parameter \(\Delta\), which is required by the horizon regularity condition. Starting from \(\Delta=1\) in the Schwarzschild limit, \(\Delta\) decreases monotonically with increasing mass and approaches zero at the endpoint of the branch. The condition \(\Delta\ge 0\) therefore determines the domain of regular scalarized solutions.

\begin{figure}[ht]
\centering
\subfigure[$S$ vs $ M$]{
\includegraphics[width=0.4\textwidth]{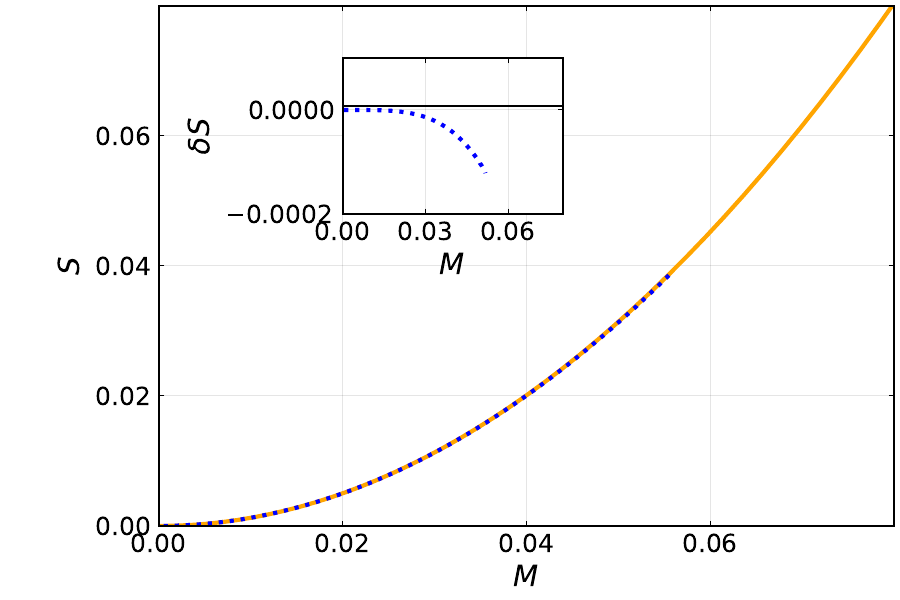}}
\hfill%
\subfigure[$\delta F $ vs $ T_H$]{\includegraphics[width=0.4\textwidth]{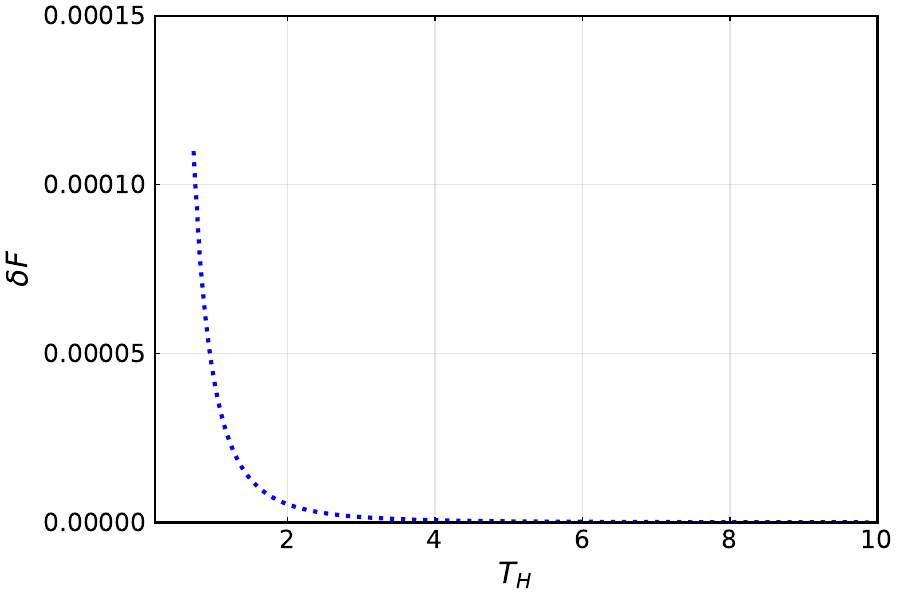}}
\caption{ Entropy $S$ and difference of free energy $\delta F$ with $\zeta_3(\phi)=\alpha\phi^4$.}\label{fig9}
\end{figure}

The corresponding thermodynamic behavior is displayed in Fig.~\ref{fig9}. The entropy \(S\) increases monotonically with \(M\), as shown in Fig.~\ref{fig9}(a). More importantly, the free-energy difference
\[
\delta F=F_{\rm scal}-F_{\rm SBH}
\]
remains non-negative over the whole temperature range, as shown in Fig.~\ref{fig9}(b). Therefore, the free energy of the scalarized black hole is never lower than that of the Schwarzschild black hole at the same temperature. In the canonical ensemble, this means that the Schwarzschild black hole is always thermodynamically preferred, while the scalarized solutions can only be regarded as metastable configurations. Accordingly, no thermodynamic phase transition is expected between the Schwarzschild and scalarized black holes for the quartic coupling.

\end{document}